\documentclass{ifacconf}

\usepackage{graphicx}      
\usepackage{natbib}        
\usepackage{amsmath}
\usepackage{amsfonts}
\usepackage{bm}
\usepackage{subcaption}
\usepackage{xcolor}

\makeatletter
\def\blfootnote{\gdef\@thefnmark{}\@footnotetext}
\makeatother

\usepackage{arydshln}
\newdimen\dashlinedash \dashlinedash2pt
\newdimen\dashlinegap \dashlinegap2pt

\begin{document}
\newcommand{\reals}{\mathbb{R}}
\newcommand{\expectation}{\mathbf{E}}
\newcommand{\expectationp}[2][]{\mathbf{E} \ifx #1 \undefined \else _{#1} \fi \left[#2\right]}
\newcommand{\variance}{\mathbf{V}}
\newcommand{\variancep}[2][]{\mathbf{V} \ifx #1 \undefined \else _{#1} \fi \left[#2\right]}
\newcommand{\dataset}{\mathcal{D}}
\newcommand{\minibatch}{\mathcal{B}}
\newcommand{\kullb}[2]{D_{\text{KL}} \left( {#1} \, \lVert \, {#2} \right)}
\newcommand{\elbo}[1]{\mathcal{L} \left( {#1} \right)}
\newcommand{\condbar}{\,|\,}  
\newcommand{\curlyb}[1]{\left\{{#1}\right\}}
\newcommand{\smallplus}{\texttt{+}}

\newcommand{\normaldist}{\mathcal{N}}
\newcommand{\lognormaldist}{\log\mathcal{N}}
\newcommand{\gammadist}{\Gamma}

\newcommand{\doo}[1]{\textup{do}(#1)}

\newcommand{\graph}{\mathcal{G}}
\newcommand{\parents}[2]{\textup{Pa}_{#1}(#2)}

\begin{frontmatter}

\title{Adjustment formulas for learning causal steady-state models from closed-loop operational data}

\author[]{Kristian Løvland$^{ *, ** }$}
\author[]{Bjarne Grimstad$^{ *, ** }$} 
\author[]{Lars Struen Imsland$^{ * }$}

\address[NTNU]{Norwegian University of Science and Technology, Trondheim, Norway, (e-mail: \{kristian.lovland, lars.imsland\}@ntnu.no)}
\address[SolSee]{Solution Seeker AS, Oslo, Norway, (e-mail: bjarne.grimstad@solutionseeker.no)}



\begin{abstract}                
Steady-state models which have been learned from historical operational data may be unfit for model-based optimization unless correlations in the training data which are introduced by control are accounted for. Using recent results from work on structural dynamical causal models, we derive a formula for adjusting for this \textit{control confounding}, enabling the estimation of a causal steady-state model from closed-loop steady-state data. The formula assumes that the available data have been gathered under some fixed control law. It works by estimating and taking into account the disturbance which the controller is trying to counteract, and enables learning from data gathered under both feedforward and feedback control.
\end{abstract}

\begin{keyword}
Learning for control, causal modelling, steady-state modelling, closed-loop identification
\end{keyword}

\end{frontmatter}

\section{Introduction}
\blfootnote{\textcolor{gray}{This work has been submitted to IFAC for possible publication}}
Data-driven modelling is increasingly being used in practical settings, including the modelling of industrial processes \citep{yin2014data,sun2021survey,Jiang2021}. However, in many cases, system measurements are a limited resource and may be infrequent due to their being impractical and expensive to gather. This is one of many reasons why one may want to use \textit{steady-state} models, which disregard transient dynamics, and only model the state of the system after it has stabilized at a constant value \citep{kadlec2009data, yin2014data, krishnamoorthy_novel_2019}. In addition to motivating the use of steady-state models, information scarcity also necessitates an efficient use of data, motivating learning from historical, \textit{operational data}, as opposed to data gathered specifically for modelling purposes. In general, and also in the real-world use case which motivated this work, different operational datasets may have been gathered under a variety of control strategies.

In this work, we examine the applicability of \textit{data-driven steady-state models} to model-based optimization, focusing on the cases where the only available data is historical operational data. We demonstrate that if the operational data are gathered from a system which is being subjected to some form of disturbance rejecting control which counteracts a slowly-varying disturbance, steady-state models learned from the data can turn out useless for optimization purposes. We explain this phenomenon using ideas from causal modelling. We also show that knowledge about the nature of the disturbances can be exploited to counteract this effect, resulting in steady-state models which \textit{are} applicable to model-based optimization.

For practitioners working with the application of models for optimization, our main result should be a natural one, since it suggests that structural causal knowledge about a system being modelled can and should be exploited to ensure satisfactory model prediction performance. This is, for instance, a principle in chemical process optimization, where structural model mismatch is a common reason for inadequate model performance \citep{quelhas2013common}.


The problem of learning steady-state models from closed-loop data is related to the problem of \textit{closed-loop system identification}, see e.g. \citet{van_den_hof_closed-loop_1998}. However, the focus of the two problems are not the same. In closed-loop system identification, the main concern is usually to identify the transient behaviour of the system in open-loop, while identifying closed-loop characteristics is of lesser importance since the control deviation due to steady-state model error will likely be accounted for by integral effect in the closed-loop controller \citep{ljung_system_2017}. In the setting we are considering, the on-line implementation of a closed-loop controller is not necessarily an option due to the possibility of low and inconsistent measurement frequency.

In the following, we will analyze the steady-state model learning problem from a \textit{causal} point of view. We emphasize that we are talking about causality as discussed in the field of causal modelling, which is concerned with modelling of and reasoning about the effects of interventions. The term should not be confused with the term as it is used in control theory and signal processing, where it indicates that a system only depends on its past and present inputs. Furthermore, the term ``causal modelling" is itself ambiguous, since it may both denote the modelling and learning of systems where the causal structure is known, and the act of learning the causal structure of a system. The problem discussed here is a case of the former. Another potential source of confusion for readers accustomed to causal modelling literature is the use of the term ``identification". In causal modelling, the term is often used to denote the process where one identifies causal structure along with expressions which enables statistical estimation of said causal effects (so-called estimands). Here, we avoid using the term whenever possible.

In Section \ref{sec:steady-state-modelling}, we define the problem of steady-state modelling. In Section \ref{sec:control-counfounding} we show using a numerical example that steady-state data from a system being controlled can contain biases due to slowly time-varying disturbances being counteracted by control. In Section \ref{sec:adjustment} we use results from causal inference to derive a method for adjusting for the biases introduced by feedforward and feedback mechanisms. In Section \ref{sec:adjustment-formula-application} we apply the adjustment formula to the numerical example from Section \ref{sec:control-counfounding}. We finish with some concluding remarks and comments concerning future work on tackling the \textit{control confounding} problem in Section \ref{sec:conclusion}.

\section{Steady-state modelling}
\label{sec:steady-state-modelling}

Consider a continuous-time dynamical system on the form
\begin{align}
    \dot{x} & = f(x, u, w) \label{eq:transition-function} \\
    y & = g(x) \label{eq:emission-function}
\end{align}
where $x$ is the system state, $u$ is the control input, $y$ is the noise-free system output, and $w$ is a process disturbance which may or may not be observed.


\begin{assum}
\label{assum:unique-steady-state}
We assume that for each pair $(u, w)$ of constant control inputs and disturbances, the system is globally asymptotically stable such that $x$ converges towards a steady-state given by the unique solution of $f(x, u, w) = 0$, regardless of its initial condition. We can define the unique steady-state resulting from the constant pair $(u, w)$ to be a function, which we denote $x_{\textup{ss}} = f_{\textup{ss}}(u, w)$.
\end{assum}

This, in turn, implies that each pair of constant control input and disturbance gives rise to a corresponding output value $y = g(f_{\textup{ss}}(u, w)) \label{eq:unique-output}$. When $w$ is a stochastic variable, the resulting steady-state $y$ also becomes a stochastic variable, which for a given $u$ is drawn from the conditional distribution $y \sim p(y \condbar u)$.

To enable the description of learning problems involving steady-state models and data, it is useful to define the term more precisely.
\begin{defn}
We define a \textit{steady-state period of tolerance $\varepsilon > 0$} to be an interval $[t^0, t^f]$, which satisfies
\begin{equation}
    \| x - x(t^0) \|_{\mathcal{L}^{\infty}(t^0, t^f)} < \varepsilon
\end{equation}
where the sup-norm $\| x \|_{\mathcal{L}^{\infty}(t^0, t^f)}$ denotes the supremum (i.e. the lowest upper bound) of $|x|$ over the interval $[t^0, t^f]$.
%
\end{defn}

\begin{assum}
The control input $u$ and process disturbance $w$ give rise to a state trajectory $x$ for which we can build a set $\{[t^0_i, t^f_i] \}_{i=1}^n$ of steady-state periods of tolerance $\varepsilon > 0$.
\end{assum}

It is natural think of this set as consisting of non-overlapping intervals of some minimum length, typically not covering all of the interval $[t_0^0, t_n^f]$ (since transient periods will be excluded from this steady-state dataset). Such a set can for instance be constructed when $w$ is varying slowly and $u$ is constant for one or more periods in time whose lengths exceed the settling time of the open-loop system dynamics. If we can construct such a set, we can extract a representative set $\mathcal{T} = \{t_i\}_{i=1}^n$ of points in time containing a representative time $t_i \in [t_i^0, t_i^f]$ for each interval. For each of these points, we consider the system to be operating at steady-state (since it is situated in a steady-state period in time). Using these time points, we can define a dataset $\dataset = \{ (u_t, y_t) \}_{t \in \mathcal{T}}$ consisting of input and output measurements considered to be taken at steady-state.

This dataset can be interpreted as a sample from the distribution $p(y \condbar u)$ at steady-state, which means it can be used to learn a steady-state model $\hat{p}(y \condbar u)$. For instance, a model could be fit to the data by minimizing mean squared error over $\dataset$. One could imagine that having access to an estimate of this steady-state distribution would enable model-based optimization of objectives containing $u$ and $y$. However, this is not necessarily the case.

\section{Biases arising from control}
\label{sec:control-counfounding}
\subsubsection{Example.}
We are given steady-state data from the system
\begin{align}
    \dot{x} & = -3x + u + w \label{eq:example-system-1}\\
    y & = 2 x \label{eq:example-system-2}
\end{align}
where the disturbance term is given by $w(t) = 2\sin(0.001t)$. We consider three different choices of control input. The choices are: 1) $u$ is a randomly chosen piecewise constant open-loop control input, 2) $u$ is given by a feedforward control mechanism which uses a measurement of the disturbance term to keep $y$ at a piecewise constant reference value $y_r$ and 3) $u$ is given by a P controller trying to keep $y$ at a piecewise constant reference value $y_r$. The implemented controllers do not actually bring $y$ to $y_r$ at steady-state due to bias and lack of integral action, respectively.

\begin{figure*}[bt]
    \centering
    \begin{subfigure}[b]{0.48\textwidth}
        \centering
        \includegraphics[width=\columnwidth]{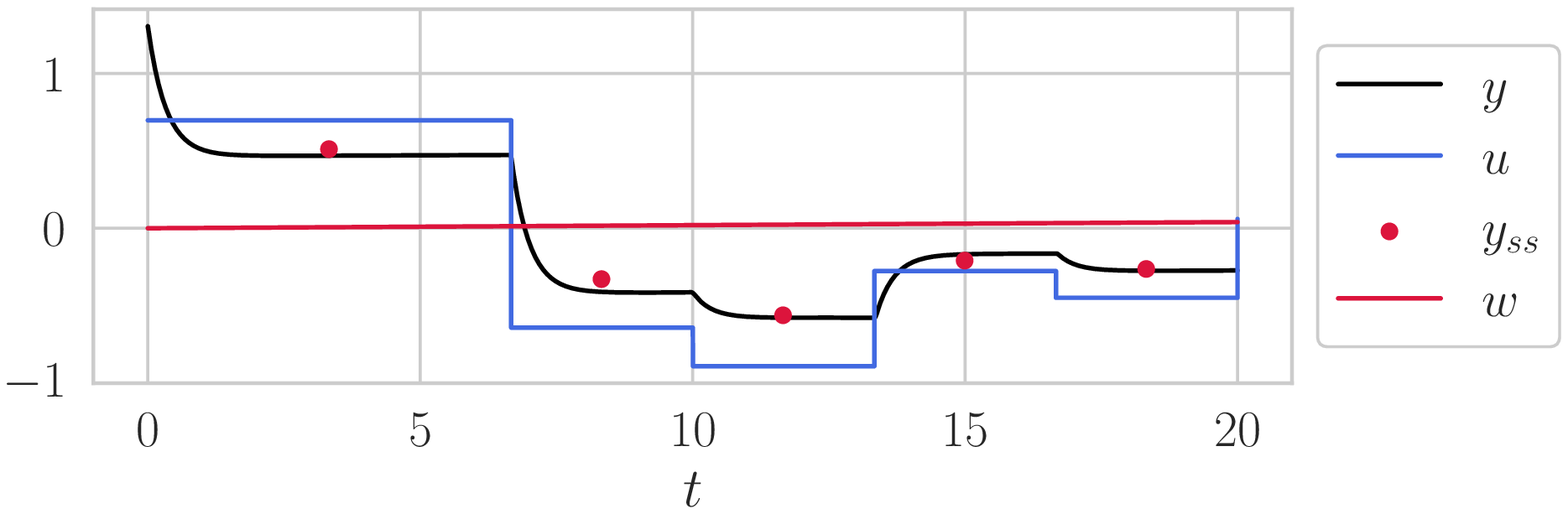}
        \caption{Fast dynamics, open-loop control}
        \label{fig:fast-open-loop-dynamics}
    \end{subfigure}
    \begin{subfigure}[b]{0.48\textwidth}
        \centering
        \includegraphics[width=\columnwidth]{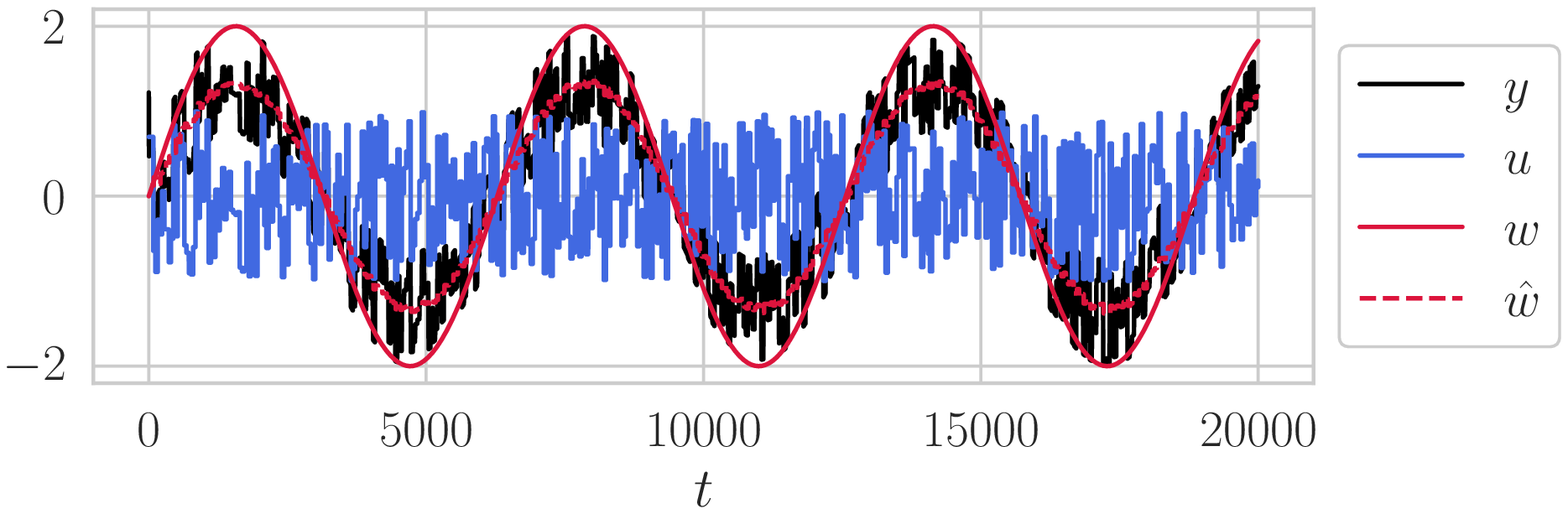}
        \caption{Slow dynamics, open-loop control}
        \label{fig:slow-open-loop-dynamics}
    \end{subfigure}

    \begin{subfigure}[b]{0.48\textwidth}
        \centering
        \includegraphics[width=\columnwidth]{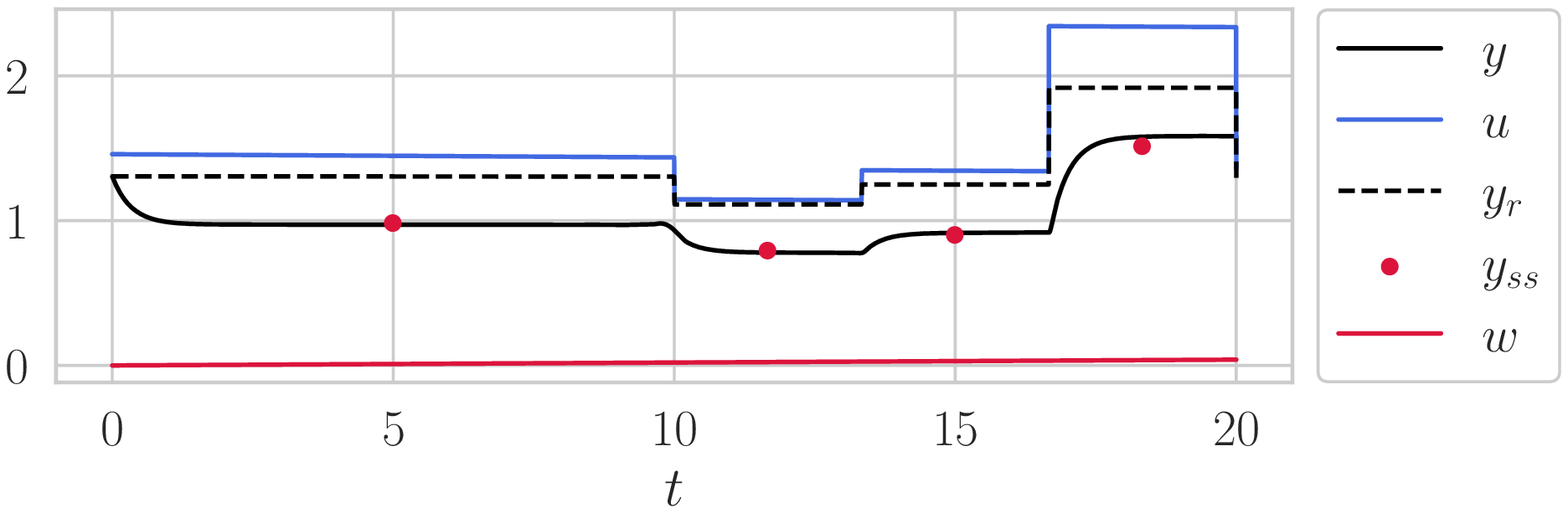}
        \caption{Fast dynamics, feedforward control}
        \label{fig:fast-feedforward-dynamics}
    \end{subfigure}
    \begin{subfigure}[b]{0.48\textwidth}
        \centering
        \includegraphics[width=\columnwidth]{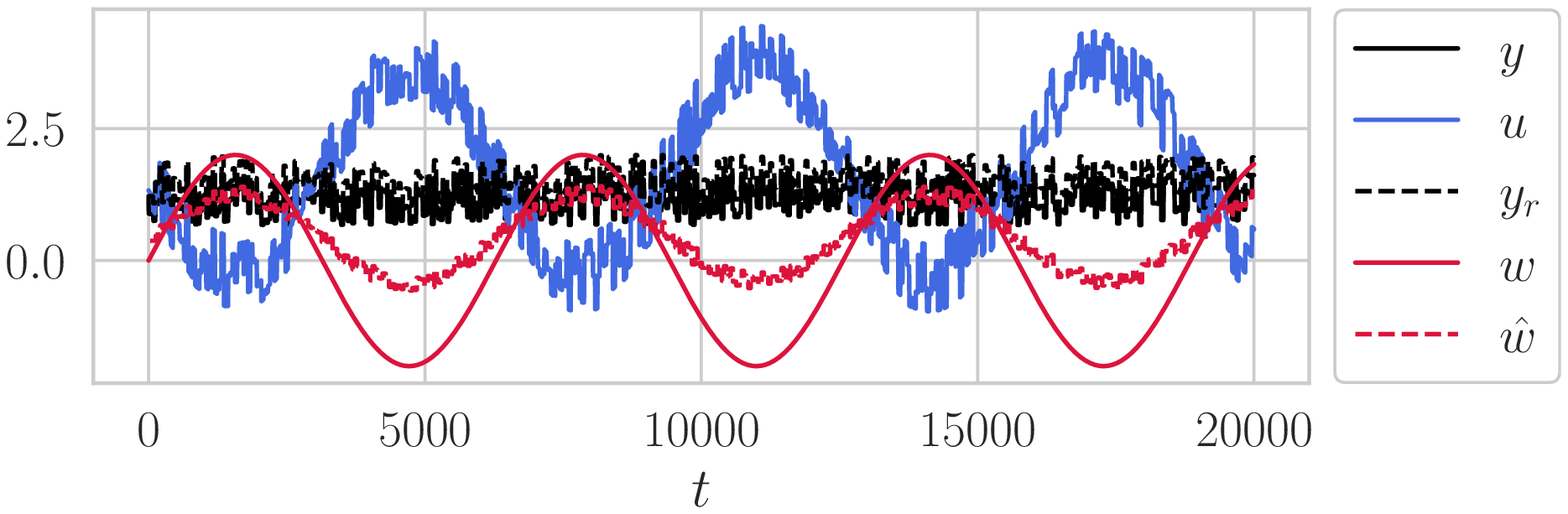}
        \caption{Slow dynamics, feedforward control}
        \label{fig:slow-feedforward-dynamics}
    \end{subfigure}
    
    \begin{subfigure}[b]{0.48\textwidth}
        \centering
        \includegraphics[width=\columnwidth]{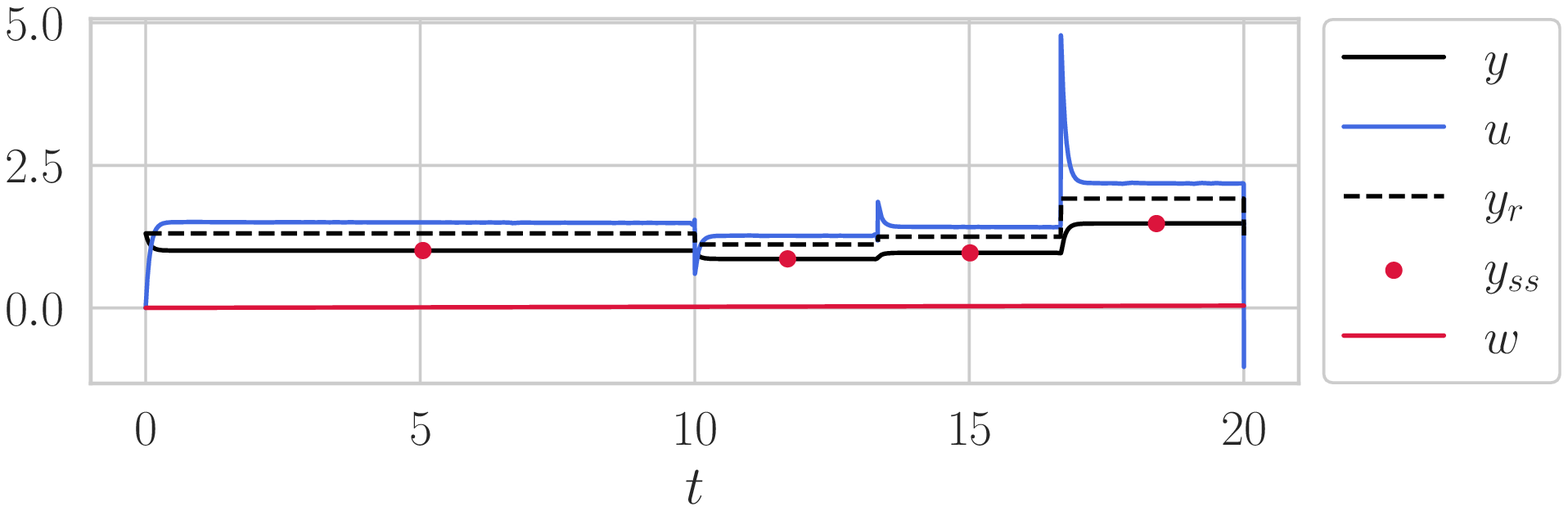}
        \caption{Fast dynamics, feedback control}
        \label{fig:fast-feedback-dynamics}
    \end{subfigure}
    \begin{subfigure}[b]{0.48\textwidth}
        \centering
        \includegraphics[width=\columnwidth]{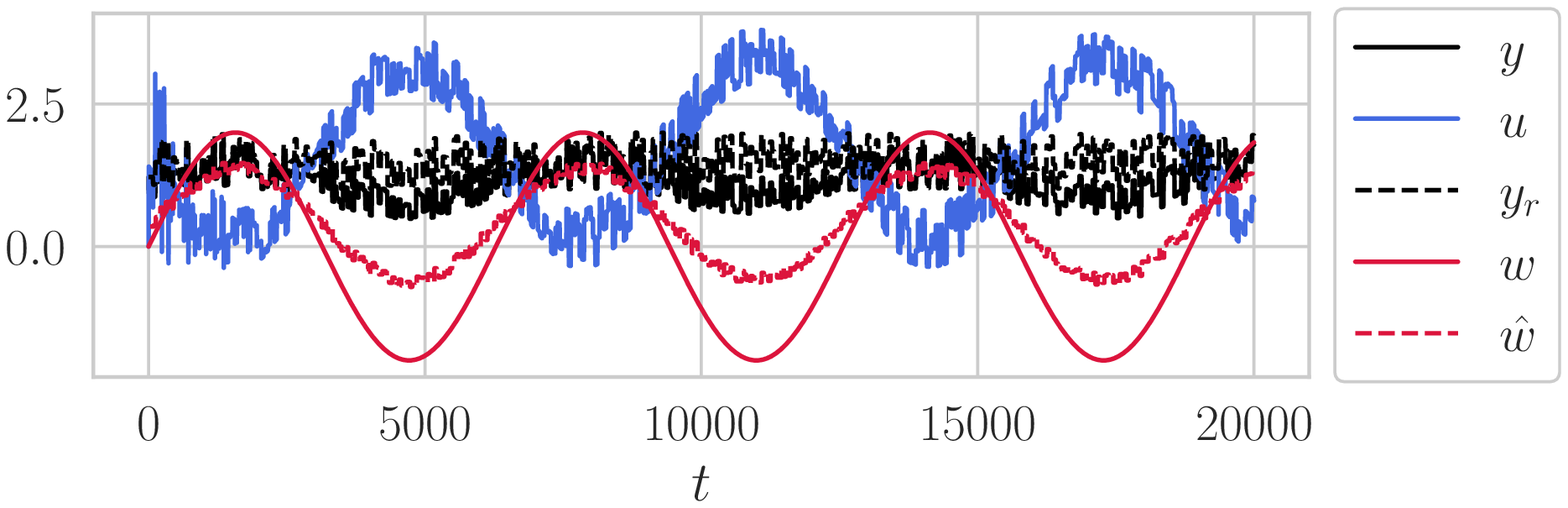}
        \caption{Slow dynamics, feedback control}
        \label{fig:slow-feedback-dynamics}
    \end{subfigure}
    \caption{Example system operating under three different types of control. The computation of the estimated disturbance $\hat w$ is explained in Section 5.}
    \label{fig:feedback-dynamics}
\end{figure*}

The dynamics of the system are significantly faster than those of the process disturbance. Thus, for all of the three cases mentioned above, the system can be interpreted as operating at a series of different steady-states. An example of this piecewise steady-state behaviour is shown in Figure \ref{fig:fast-open-loop-dynamics}, which illustrates the fast dynamics of the system as well as identified steady-state periods. The slow dynamics of the system is illustrated in Figure \ref{fig:slow-open-loop-dynamics}. Similar illustrations showing the same system being subjected to feedforward and feedback control are shown in Figures \ref{fig:fast-feedforward-dynamics}, \ref{fig:slow-feedforward-dynamics}, \ref{fig:fast-feedback-dynamics} and  \ref{fig:slow-feedback-dynamics}.

\begin{figure*}[bt]
    \centering
    \begin{subfigure}[b]{0.32\textwidth}
            \centering
            \includegraphics[width=\textwidth]{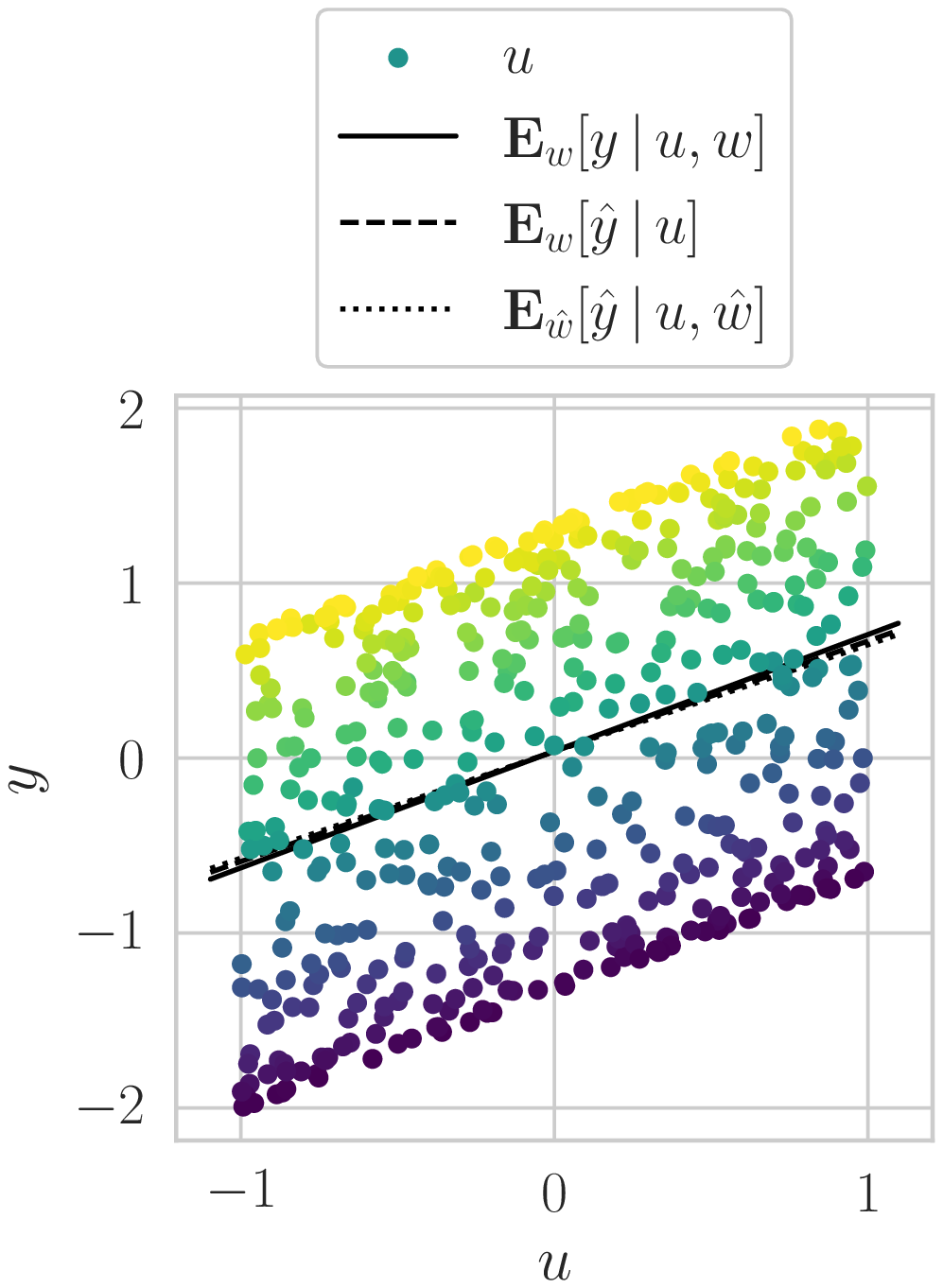}
            \caption{Open-loop}
            \label{fig:uy-scatterplot-open-loop}
        \end{subfigure}
    \begin{subfigure}[b]{0.32\textwidth}
            \centering
            \includegraphics[width=\textwidth]{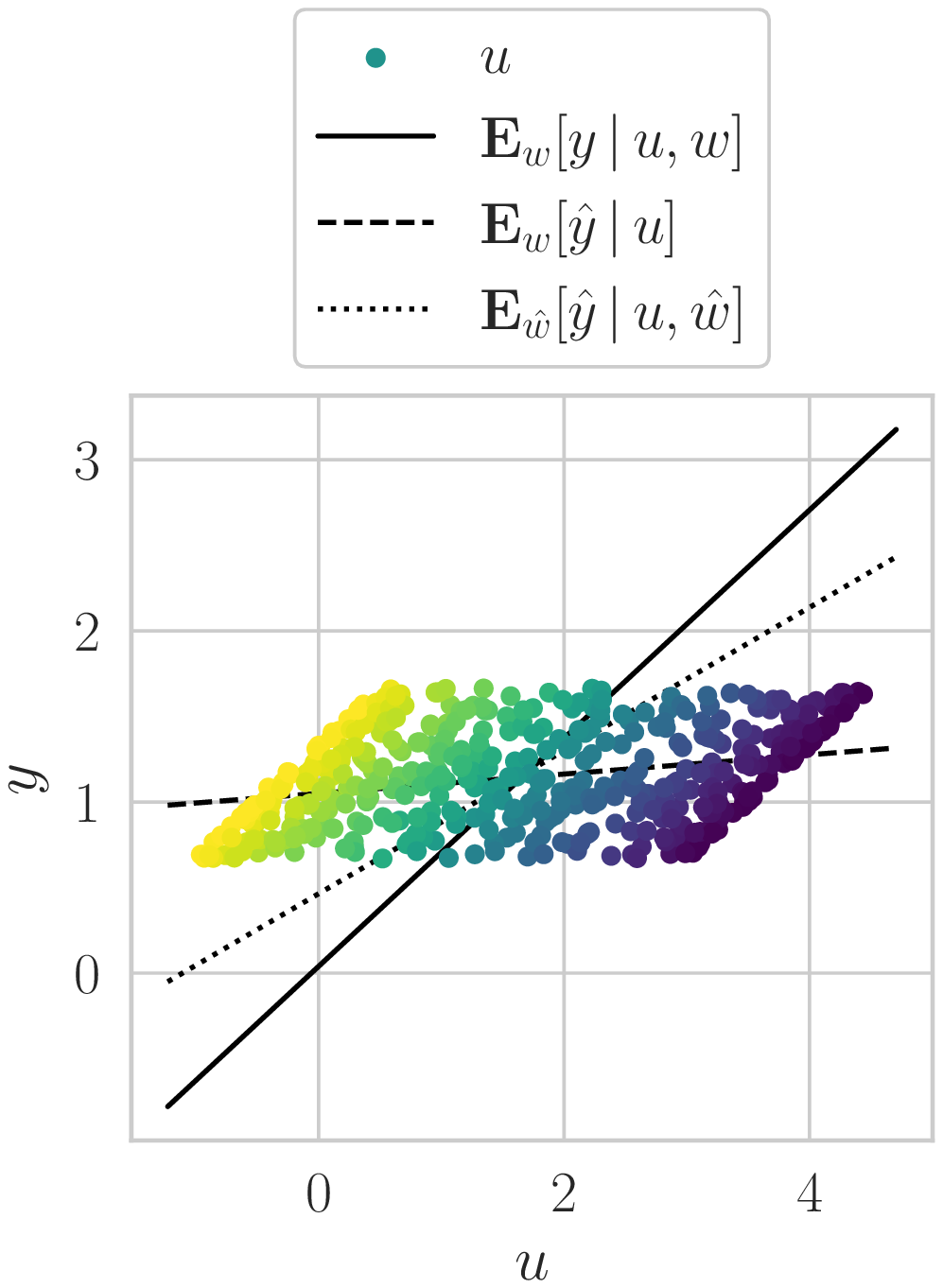}
            \caption{Feedforward control}
            \label{fig:uy-scatterplot-feedforward}
        \end{subfigure}
    \begin{subfigure}[b]{0.32\textwidth}
            \centering
            \includegraphics[width=\textwidth]{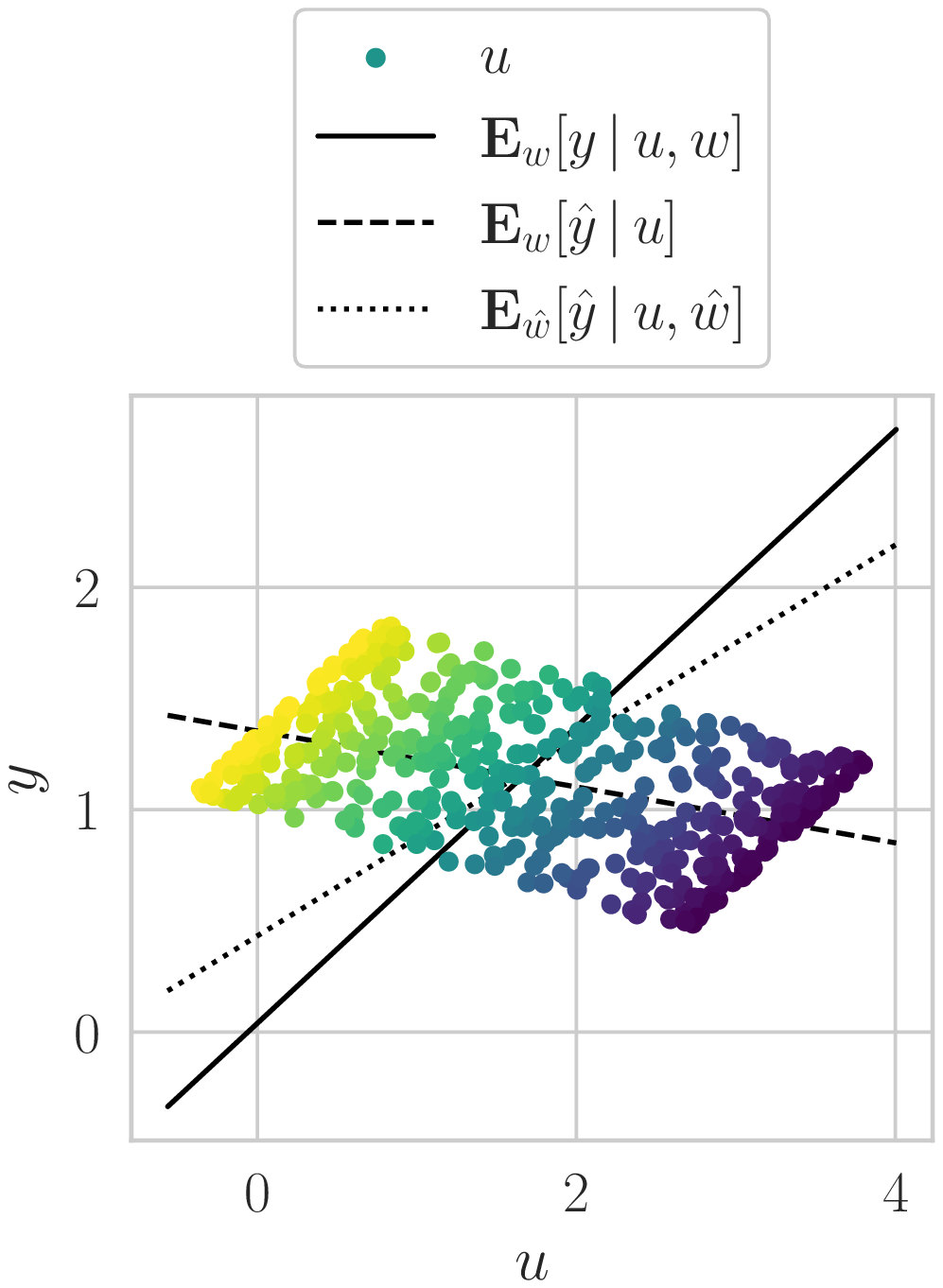}
            \caption{Feedback control}
            \label{fig:uy-scatterplot-feedback}
        \end{subfigure}
    \caption{Steady-state relation between $u$ and $y$ under different kinds of control strategies. Points in the scatterplot show steady-state measurements. The color of the points indicate the value of the process disturbance which was present at the steady-state each data point was collected (ranging from -2 to 2). Solid lines show the true steady-state relation between $u$ and $y$, dashed lines show linear models fitted using least squares without any adjustment, and dotted lines show linear models which use the derived adjustment formula. All of the models are shown with $w$ or $\hat{w}$ being at their mean, respectively. The computation of the estimated disturbance $\hat w$ is explained in Section 5.}
    \label{fig:uy-scatterplots}
\end{figure*}

Figure \ref{fig:uy-scatterplots} shows the relation between control input and resulting steady-state output for a dataset of detected steady-states. The dataset has been built by first iterating through all time indices and detecting steady-state periods of length 3 which have tolerance $\varepsilon = 0.05$, and then piecing them together afterwards if they are overlapping. A single data point containing the average over the steady-state period is added to the dataset for each of the identified steady-state periods.

We consider the problem of fitting a linear steady-state model
\begin{equation}
    \hat{y} = \hat{c}_0 + \hat{c}_1 u \label{eq:non-causal-linear-model}
\end{equation}
to the steady-state data from the system. Figure \ref{fig:uy-scatterplots} shows least squares fits for data gathered under the three types of control. In the case of open-loop control, the parameter estimate looks good. In the cases of feedforward and feedback control however, the error of the estimates are significant. In the feedback control case, the estimated effect on $y$ of changes in $u$ even has the wrong sign. Hence, a model $\hat{p}(y \condbar u)$ fitted to these data would be completely useless for model-based optimization.

\section{Adjusting for the biases using causal insights}
\label{sec:adjustment}
The phenomenon shown in the previous section can be explained using tools from causal inference. For a thorough introduction to the field, see \citet{pearl_causality_2009}. Of greatest importance to us is the distinction it introduces between observing and intervening, which is formalized through the so-called do-operator. Briefly put, this operator defines the probability $p(A \condbar \doo{B})$, which in general is not the same as $p(A \condbar B)$. An example is the system where $A$ is a binary variable indicating whether or not it rains and $B$ is a binary variable indicating whether a given person has opened their umbrella. Observing  a person closing their umbrella would probably change one's beliefs about whether it is currently raining, while going physically intervening and closing the umbrella would probably not. The do-operator formalizes this intuitive notion, enabling one to state the classical mantra that ``correlation is not causation" in a rigorous way.

From the viewpoint of causal modelling, the behaviour observed in the above example is to be expected. The \textit{observational} distribution $p(y \condbar u)$ contains correlations which are present in the dataset, and the model has no reason not to capture the dependencies introduced by the control law. If one wishes to use the steady-state model for model-based optimization, the probability distribution of interest should be the \textit{interventional} distribution $p(y \condbar \doo{u})$.

The simplest way to estimate interventional distributions is to intervene on the system and observe the resulting effects. This is exactly what was done in the open-loop case in the example in the previous section, which was the only setting where the data gave rise to a steady-state model which exhibited the expected (that is to say, causal) behaviour.

In situations where intervening on the system is not an option, estimating distributions containing causal relations may still be possible if the causal structure of the system being observed is known. In the following, we will exploit structural knowledge about causal relations in control systems to derive a formula which enables the estimation of $p(y \condbar \doo{u})$ from closed-loop steady-state data. In Section \ref{sec:scm}, we give an introduction to the framework of the \textit{structural causal model} (SCM), which enables high-level description of causal relations. In Section \ref{sec:sdcm}, we introduce the \textit{structural dynamical causal model} (SDCM), which generalizes the SCM to dynamical systems. In Section \ref{sec:backdoor-adjustment} we describe the \textit{backdoor-adjustment formula}, which under certain conditions can be used to estimate causal quantities from observational data. In Section \ref{sec:feedforward-adjustment} we use recent results concerning SDCMs together with the backdoor-adjustment formula to derive a formula for estimating $p(y \condbar \doo{u})$ using data which have been gathered under feedforward control. Finally, in Section \ref{sec:feedback-adjustment} we show that under certain assumptions, the same formula also holds for data from feedback controlled systems.

Sections \ref{sec:scm} and \ref{sec:sdcm} will largely follow the introduction to SCMs and SDCMs given by \citet{bongers_causal_2022}, albeit in a somewhat simplified and chronologically different manner. For a more detailed introduction, which explicitly treats the measure theoretic foundations of the SCM and the SDCM, we refer to the original source.

\subsection{Structural causal models}
\label{sec:scm}
A structural causal model $\mathcal{M}$ consists of a set of exogenous variables $\{e_j\}_{j \in \mathcal{J}} = e \sim p(e)$ and a set of endogenous variables $\{z_i\}_{i \in \mathcal{I}} = z$, whose values, for a given assignment $e$, are determined by the causal mechanism
\begin{equation}
    z = h(z, e) \label{eq:basic-scm}
\end{equation}
Assuming that the endogenous and exogenous variables are defined over suitable domains of appropriate dimension, and omitting the index sets of the elements of $z$ and $e$ from the definition for conciseness, we define the SCM as a pair consisting of a causal mechanism and an exogenous variable distribution, that is $\mathcal{M} := (h, p(e))$. Note that the endogenous variables do not enter explicitly into the definition, since they are defined implicitly through $h$.

If (\ref{eq:basic-scm}) has a unique solution for any given $e$, the function $h$ can be thought as an assignment from exogenous variables to endogenous variables. Since $e$ is a stochastic variable, the assignment also gives rise to a distribution $p(z)$, which we will denote $p_{\mathcal{M}}(z)$ to emphasize the fact that it is defined by $\mathcal{M}$.

Furthermore, any SCM $\mathcal{M}$ is equipped with an operation called perfect intervention, which models the act of intervening on states of the system. The perfect intervention $\doo{z_I = \zeta_I}$ exchanges the causal mechanism $h$ with the \textit{intervened causal mechanism} $\tilde{h}$, given by
\begin{equation}
    \tilde{h}_i(z, e) :=
    \begin{cases}
    \zeta_i, \quad & i \in I \\
    h_i(z, e), \quad & i \in \mathcal{I} \setminus I
    \end{cases}
\end{equation}
We denote the resulting \textit{intervened SCM} $\mathcal{M}(\doo{z_I = \zeta_I}) = ( \tilde{h}, p(e) )$. We may also denote it $\mathcal{M}(\doo{z_I})$ when the interpretation of $z_I$ is clear from context. If (\ref{eq:basic-scm}) is uniquely solvable for any given $e$, this operation can be thought of as overriding the assignment which in an unintervened setting would give rise to $z_I$, forcing $z_I$ to equal $\zeta_I$ instead. This operation does not need be possible to perform in practice to be valid. If the intervention is imaginable, modelling the consequences of a hypothetical intervention can still be meaningful even though it is not practically feasible.

Typically, each element $h_i$ of the causal mechanism will only depend on subsets of $z$ and $e$. The elements of $z$ and $e$ on which $h_i$ depend are called the endogenous and exogenous parents of $z_i$ (denoted $\parents{\mathcal{I}}{i}$ and $\parents{\mathcal{J}}{i}$), respectively. These dependencies can be represented by a graph, where each endogenous variable $i \in \mathcal{I}$ have parents $\parents{\mathcal{I}}{i} \cup \parents{\mathcal{J}}{i}$. The causes of the exogenous variables are not modelled, and they are instead represented by random variables (what one in a control setting would often call ``disturbance" or ``noise").

This \textit{causal graph} encodes information about the direct causal relations which are implied by $\mathcal{M}$, and it can be a useful tool for discussing independence properties (both statistical and causal) while making quite generic assumptions about the functional forms of $h$ and $p(e)$. The most important results, including the ones we will use, rely on the concept of \textit{d-separation}, see e.g. \citet{koller_probabilistic_2009}.

\subsection{Structural dynamical causal models}
\label{sec:sdcm}
The \textit{structural dynamical causal model} (SDCM), introduced in \cite{bongers_causal_2022}, extends the causal semantics of the SCM to dynamical systems. The extension is done by writing the system dynamics on the form
\begin{equation}
    z = h(z, \dot{z}, \dots, z^{(n)}, e)
\end{equation}
If a process $z$, which is assumed to be $n$ times differentiable, satisfies the above equation for all $t$, it is called a solution to the SDCM, and the function $h$ can be thought of as an assignment. The interpretation of this assignment function may not be as clear as it was for the SCM. One justification for writing the dynamics on this implicit form is that it makes an extension of the perfect intervention to the dynamical case notationally simple. For an SDCM, the perfect intervention $\doo{z_I = \zeta_I}$ is defined as
\begin{equation}
    \tilde{h}_i(z, \dot{z}, \dots, z^{(n)}, e) :=
    \begin{cases}
    \zeta_i,  & i \in I \\
    h_i(z, \dot{z}, \dots, z^{(n)}, e),  & i \in \mathcal{I} \setminus I
    \end{cases}
\end{equation}
Intuitively, an SDCM is an SCM where the variables are dynamical processes. Then, we can think of the perfect intervention $\doo{z_I = \zeta_I}$ as an operation where one keeps the set $z_I$ of time-varying variables fixed at $\zeta_I$ forever, instead of letting them evolve naturally over time. This interpretation will be the basis for the causal steady-state models which will be derived later.

An important result regarding SDCMs, first stated in \citet{bongers_causal_2022}, concerns the \textit{equilibration} operation, which is performed by setting all higher-order derivatives to zero and solving the resulting equation $z = h(z, 0, \dots, 0, e)$ whose solution is the equilibrium resulting from the exogenous variable taking the constant value $e$. The result states that for a certain class of well-behaved SDCMs (called \textit{steady} SDCMs), equilibration and intervention commutes. Thus, if one only cares about the steady-state resulting from an intervention, one only needs to consider the equilibrated SDCM, which is itself an SCM.

\subsection{Adjustment criterion for acyclic models}
\label{sec:backdoor-adjustment}
The so-called \textit{backdoor-adjustment} (see e.g. \citep{pearl_causality_2009}) is a key result in causal modelling. It consists of a criterion and a formula which together enable the estimation of causal quantities from purely observational data. The insight which underlies backdoor-adjustment is that non-causal correlations between variables are often due to common, confounding causes (often simply called \textit{confounders}). If an SCM has an acyclic graph and independent exogenous variables, the existence of confounders between two variables in a given SCM can be checked using the graph of the SCM and its d-separation properties. Specifically, the criterion gives sufficient conditions for when a set of variables, concatenated into the vector $r$, can be used to adjust for confounders when the goal is to estimate the causal effect of the variables $u$ on $y$. The criterion requires that 1) No element of $r$ is a descendant of $u$ and 2) The elements of $r$ d-separates all paths from $u$ to $y$ that contains an arrow pointing towards $u$ (i.e. all the ``backdoor" paths). A set of variables which satisfies both of these criteria is called \textit{sufficient for adjustment}.

\subsubsection{Example.}
\begin{figure*}[bt]
    \centering
    \begin{subfigure}[b]{0.45\textwidth}
            \centering
            \includegraphics[width=6.1cm]{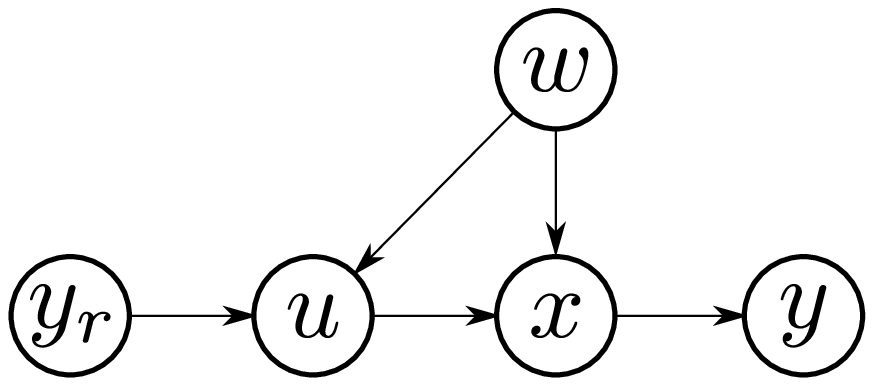}    
            \caption{Feedforward control} 
            \label{fig:feedforward-scm}
        \end{subfigure}
    \begin{subfigure}[b]{0.45\textwidth}
            \centering
            \includegraphics[width=6.1cm]{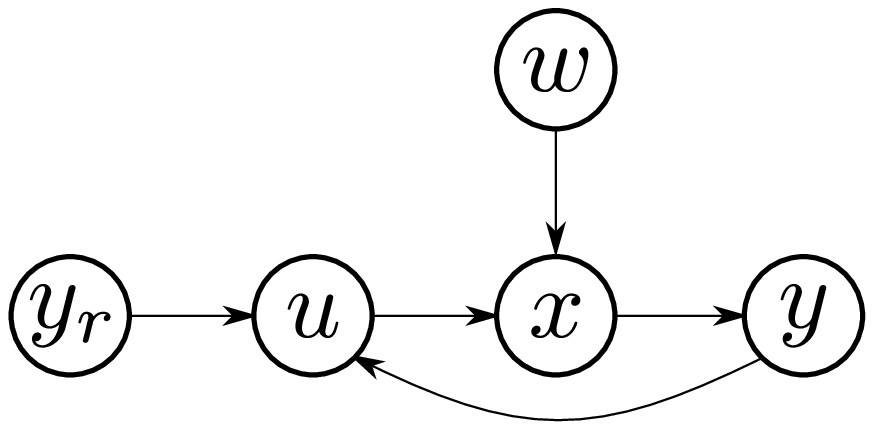}    
            \caption{Feedback control}
            \label{fig:feedback-scm}
        \end{subfigure}
    \caption{Structural causal models for the steady-state of a dynamic system operating under control}
    \label{fig:scm}
\end{figure*}

The criterion can be illustrated using the SCM shown in Figure \ref{fig:feedforward-scm}, for which we want to learn $p(y \condbar \doo{u})$ from observational data. Consider first the candidate set $r = \emptyset$. Under this set of adjustment variables, the path $u \leftarrow w \rightarrow x \rightarrow y$ is an unblocked backdoor path, meaning 1) does not hold. Thus, the empty set is not sufficient for adjustment. The set consisting of $w$ however, \textit{is} sufficient for adjustment. To see this, note first that the backdoor path $u \leftarrow w \rightarrow x \rightarrow y$ becomes d-blocked when $w$ is observed. The path $u \rightarrow x \rightarrow y$ is still active, but since it contains no arrows into $u$, it is not a backdoor path. Thus, since $w$ is not a descendant of $u$, both 1) and 2) holds, and $w$ is sufficient for adjustment.

If the criterion holds, one has that
\begin{align}
    p(y \condbar \doo{u}) & = \int p(y \condbar u, r) p(r) dr \\
    & = \expectation_{r \sim p(r)}\left[ p(y \condbar u, r) \right]
\end{align}
This result is significant in that it enables estimation of interventional distributions without actually intervening on the system, provided that the assumptions about causal independencies which are made by the SCM are correct and sufficiently informative.

\subsection{Feedforward control}
\label{sec:feedforward-adjustment}
Consider a setting where the steady-state data have been gathered with the system being subject to feedforward control, such that
\begin{equation}
    u = k_{\textup{ff}}(y_r, w) \label{eq:feedforward-control}
\end{equation}
The system of ODEs given by coupling (\ref{eq:transition-function}), (\ref{eq:feedforward-control}) and (\ref{eq:emission-function}) describes the dynamical behaviour of the system. However, our objective is to model what happens to the steady-state of this system when $u$ is set to some constant value.

As shown in Appendix \ref{sec:feedforward-scm-derivation}, the system can be rewritten as an SDCM. The control action of setting $u$ to a constant value can then be expressed as the intervention $\doo{u}$. Since we only care about the steady-state resulting from this intervention, we can simplify the modelling problem by equilibrating the SDCM, as shown in Appendix \ref{sec:feedforward-scm-derivation}. This results in an SCM, which we denote $\mathcal{M}_{\textup{ff}}$, and whose graph is shown in Figure \ref{fig:feedforward-scm}.
\begin{thm}
\label{thm:feedforward-adjustment}
Let $\mathcal{M}_{\textup{ff}} = ((f_{\textup{ss}}, k_{\textup{ff}}, g),\, p(y_r, w) )$. Then,
\begin{align}
    p_{\mathcal{M}_{\textup{ff}}}(y \condbar \doo{u}) & = \int p_{\mathcal{M}_{\textup{ff}}}(y \condbar u, w) p_{\mathcal{M}_{\textup{ff}}}(w) dw \label{eq:feedforward-adjustment}
\end{align}
\end{thm}
\begin{pf}
Since the graph of $\mathcal{M}_{\textup{ff}}$ is acyclic, the causal effect of $u$ on $y$ is a candidate for being estimated using the backdoor criterion. As shown in the above example, the variable $w$ is sufficient for adjustment. Applying the backdoor-adjustment formula, we get (\ref{eq:feedforward-adjustment}). \qed
\end{pf}
In other words, it is possible to estimate the effect of intervening on $u$ using observational steady-state data gathered under feedforward control, through samples from $p_{\mathcal{M}_{\textup{ff}}}(y \condbar u, w)$ and $p_{\mathcal{M}_{\textup{ff}}}(w)$. That is, if one is willing to assume that $\mathcal{M}_{\textup{ff}}$ accurately describes the causal structure of the data-generating process.

The feedforward adjustment formula introduces a new dependence in the distribution over the system output, which is now given by $p_{\mathcal{M}_{\textup{ff}}}(y \condbar u, w)$. Intuitively, this distribution takes into account the fact that the effect control input has on system output depends on the process disturbance it will have to counteract.  To be able to evaluate the integral one also needs to evaluate $p_{\mathcal{M}_{\textup{ff}}}(w)$. This distribution does not need to come from direct observations, but can also come from other, indirect types of estimates.

\subsection{Feedback control}
\label{sec:feedback-adjustment}
Consider now a setting where the data have been gathered with the system being subject to a feedback control law on the form
\begin{equation}
    u = k(y_r, y)
\end{equation}
Assuming that the controller manages to make the closed-loop system converge to some steady-state for each possible $y_r$, the steady-state system behaviour can be derived in a very similar manner as in the feedforward case, resulting in the graph shown in Figure \ref{fig:feedback-scm}. Furthermore, under stricter assumptions on the controller, one can show that the adjustment formula from before also holds for this feedback case.

\begin{thm}
Let $\mathcal{M}_{\textup{fb}} = ((f_{\textup{ss}}, k_{\textup{fb}}, g),\, p(y_r, w) )$. Assume that the feedback controller $k_{\textup{fb}}$ is able to bring the system to a steady-state $x_{\textup{ss}} = k_{\textup{ff}}'(y_r, w)$ for any combination of reference and disturbance. Then,
\begin{align}
    p_{\mathcal{M}_{\textup{fb}}}(y \condbar \doo{u}) & = \int p_{\mathcal{M}_{\textup{fb}}}(y \condbar u, w) p_{\mathcal{M}_{\textup{fb}}}(w) dw \label{eq:feedback-adjustment}
\end{align}
\end{thm}
\begin{pf}
Define $\mathcal{M}_{\textup{ff}}' = ((f_{\textup{ss}}, k_{\textup{ff}}', g),\, p(y_r, w) )$. By construction, the causal mechanisms of $\mathcal{M}_{\textup{fb}}$ and $\mathcal{M}_{\textup{ff}}'$ return the same $(x, u, y)$ for any choice of exogenous variables $(y_r, w)$. Since the two SCMs have the same exogenous variable distribution $p(y_r, w)$, they are then observationally equivalent, meaning
\begin{equation}
p_{\mathcal{M}_{\textup{fb}}}(x, u, y, y_r, w) = p_{\mathcal{M}_{\textup{ff}}'}(x, u, y, y_r, w)    
\end{equation}
In particular, $p_{\mathcal{M}_{\textup{fb}}}(y \condbar u, w) = p_{\mathcal{M}_{\textup{ff}}'}(y \condbar u, w)$ and $p_{\mathcal{M}_{\textup{fb}}}(w) = p_{\mathcal{M}_{\textup{ff}}'}(w)$. Furthermore, note that
\begin{equation}
    \mathcal{M}_{\textup{fb}}(\doo{u}) = ((f_{\textup{ss}}, u, g),\, p(y_r, w) ) = \mathcal{M}_{\textup{ff}}'(\doo{u})
\end{equation}
such that
\begin{equation}
    p_{\mathcal{M}_{\textup{fb}}}(y \condbar \doo{u}) = p_{\mathcal{M}_{\textup{ff}}'}(y \condbar \doo{u})    
\end{equation}
Finally, note that $\mathcal{M}_{\textup{ff}}'$ satisfies the conditions of Theorem \ref{thm:feedforward-adjustment}. Then,
\begin{align}
    p_{\mathcal{M}_{\textup{fb}}}(y \condbar \doo{u}) & = p_{\mathcal{M}_{\textup{ff}}'}(y \condbar \doo{u}) \\
    & = \int p_{\mathcal{M}_{\textup{ff}}'}(y \condbar u, w) p_{\mathcal{M}_{\textup{ff}}'}(w) dw \\
    & = \int p_{\mathcal{M}_{\textup{fb}}}(y \condbar u, w) p_{\mathcal{M}_{\textup{fb}}}(w) dw
\end{align}
\flushright\qed
\end{pf}

This may at first glance seem counter-intuitive. One way to interpret is as follows: For the feedback controller to keep $y$ close to $y_r$ under changing external circumstances, it in practice needs to estimate these external circumstances. In our case, the external circumstances come in the form of the disturbance $w$. Thus, at steady-state, there is no way to distinguish between a feedback controller and a feedforward controller which gives rise to the same steady-state. This fact is the key property which makes the feedforward adjustment formula applicable to steady-state data gathered from feedback controlled systems.

We stress that we make no assumptions regarding the performance of the controllers, and that as long as one is willing to assume that the controllers stay consistent for the duration of the dataset, it does not matter if they are model-based controllers having access to high-frequency output measurements, or if they are human operators which looks at a noisy estimate of the system output and changes the system control input once.

\section{Applying the adjustment formula}
\label{sec:adjustment-formula-application}
Consider again the example from Section \ref{sec:control-counfounding}. Since we want to use our steady-state model for optimization, we state the intervened SCM
\begin{align}
    x & = f_{\textup{ss}}^u u + f_{\textup{ss}}^w w \label{eq:linear-mff-1} \\
    y & = g x \label{eq:linear-mff-2} \\
    y_r & \sim \normaldist(\mu_{y_r}, \sigma_{y_r}^2) \label{eq:linear-mff-3} \\
    w & \sim \normaldist(\mu_w, \sigma_w^2 I) \label{eq:linear-mff-4}
\end{align}
which is a special case of $\mathcal{M}_{\textup{ff}}(\doo{u})$ where all variables are scalar, the causal mechanism is linear, and all exogenous variables are Gaussian. The parameters of the model $p(y \condbar \doo{u})$ under this model can be learned using the dataset from Section \ref{sec:control-counfounding}, which can be written as vectors $\bm u = [u_{t_1} \dots u_{t_n}]^T$, $\bm y = [y_{t_1} \dots y_{t_n}]^T$, which have corresponding unobserved disturbance $\bm w = [w_{t_1} \dots w_{t_n}]^T$.

We define steady-state input-output gain $c = g f_{\textup{ss}}^u$. Furthermore, we assume that $w$ is scaled such that $g f_{\textup{ss}}^w = 1$. This will result in the estimated disturbance and the actual disturbance having different scales, a phenomenon which is typical for latent variables. For the moment, assume that $c$ has a prior distribution $p(c)$. Then,
\begin{equation}
    p(c, \bm w \condbar \bm u, \bm y) \propto p(\bm y \condbar \bm u, c, \bm w) p(\bm u \condbar c, \bm w) p(c) p(\bm w)
\end{equation}
We assume that the controller, model parameters and process disturbance are completely unknown to us, which we represent by placing uninformative priors on them (i.e. we let $\sigma_u^2, \sigma_{c}^2, \sigma_w^2 \rightarrow \infty$). Thus, we disregard the terms $p(\bm u \condbar c, \bm w)$, $p(c)$ and $p(\bm w)$ from now on. We are left with the disturbance adjusted model, which is defined by
\begin{equation}
    p(\bm y \condbar \bm u, c, \bm w) = \delta(c \bm u + \bm w)
\end{equation}
where $\delta$ denotes the Dirac delta distribution. A MAP estimate of $(c, \bm w)$ can then be derived using standard methods, and is given by the solution $[ \hat{c} \quad \hat{\bm w}^T ]^T$ of the equation
\begin{equation}
    \left[
    \begin{array}{c:c}
    \bm u^T \bm u & \bm u^T \\ \hdashline
    \bm u & \bm I
    \end{array}
    \right]
    \left[
    \begin{array}{c}
    c \\ \hdashline
    \bm w
    \end{array}
    \right]
    =
    \left[
    \begin{array}{c}
    \bm u^T \bm y \\ \hdashline
    \bm y
    \end{array}
    \right]
    \label{eq:linear-gaussian-map-estimator}
\end{equation}
resulting in the causal steady-state model $\hat{y}_t = \hat{c} u_t + \hat{w}_t$.

Inserting the data from the example of Section \ref{sec:control-counfounding} into the MAP estimator of (\ref{eq:linear-gaussian-map-estimator}) results in the estimates $\hat{w}$ that are shown in Figures \ref{fig:slow-open-loop-dynamics}, \ref{fig:slow-feedforward-dynamics} and \ref{fig:slow-feedback-dynamics}. Figure \ref{fig:uy-scatterplots} shows the fitted steady-state for the case of zero disturbance. The disturbance estimation is by no means perfect, but it does help in removing a significant amount of the control confounder bias. If more knowledge about the system was available, it could be used to improve upon the assumptions made in the above derivation, for instance by trying to estimate controller parameters, or by specifying a more informative disturbance covariance matrix. Such an introduction of knowledge would likely result in a more precise model.

The phenomenon of control confounding can be seen as a version Simpson's paradox, which is a statistical phenomenon where regression performed on a whole dataset and regression performed individually on subsets of the same dataset grouped by some category give rise to drastically different model predictions \citep{simpson1951interpretation}. Figure \ref{fig:uy-scatterplot-feedback} illustrates this, by showing that regression performed on subsets of the dataset which have similar values of $w$ give positive model slopes, while regression performed on the whole dataset gives negative slope. From this viewpoint, the adjustment formula of (\ref{eq:feedforward-adjustment}) is simply performing grouping on $w$ before regression.


\section{Conclusion and further work}
\label{sec:conclusion}
We showed that steady-state models learned from operational data gathered under control can be unsuited for use in model-based optimization since they in general fail to model the interventional distribution $p(y \condbar \doo{u})$. Motivated by this, we showed that models which are suited for model-based optimization \textit{can} still be learned from observational data, as long as one is able to model said interventional distribution. Starting from a generic nonlinear ODE describing the dynamical behaviour of a system being controlled, we used results from causal modelling to derive an adjustment formula. The formula exploits knowledge about the process disturbances which the controller is trying to counteract, and enables the estimation of the steady-state $p(y \condbar \doo{u})$ from operational data. We illustrated both the problem and our proposed solution using a numerical example.

Investigating systems whose causal dependencies during operation are not accurately described by any of the graphs in Figure \ref{fig:scm} these is an interesting topic for further research. Other natural topics to investigate further is to what extent the adjustment formula can be adapted to settings with output measurement noise, controllers which have an internal state (e.g. PI controllers), more general types of process disturbance, and settings where the controller may also be time varying.


\bibliography{ifacconf}             
\appendix
\section{Derivation of equilibrated SDCM}
\label{sec:feedforward-scm-derivation}
Combining (\ref{eq:transition-function}), (\ref{eq:feedforward-control}) and (\ref{eq:emission-function}), we get the system
\begin{align}
    \dot{x} & = f(x, u, w) \\
    u & = k_{\textup{ff}}(y_r, w) \\
    y & = g(x)
\end{align}
which can be written as the SDCM
\begin{align}
    x & = x - \dot{x} + f(x, u, w) \\
    u & = k_{\textup{ff}}(y_r, w) \\
    y & = g(x) \\
    y_r, w & \textup{ exogenous}
\end{align}
Assuming that $f$ is continuous, and that the exogenous variables $y_r$ and $w$ are constant, independent and drawn from $p(y_r)$ and $p(w)$, the equilibrated system can be written as
\begin{align}
    x & = x + f(x, u, w) \label{eq:self-cycle1}\\
    u & = k_{\textup{ff}}(y_r, w) \\
    y & = g(x) \\
    y_r & \sim p(y_r) \\
    w & \sim p(w)
\end{align}
By Assumption \ref{assum:unique-steady-state}, we can write an explicit solution to (\ref{eq:self-cycle1}), removing the self-cycle and resulting in the SCM
\begin{align}
    x & = f_{\textup{ss}}(u, w) \label{eq:self-cycle2}\\
    u & = k_{\textup{ff}}(y_r, w) \\
    y & = g(x) \\
    y_r & \sim p(y_r) \\
    w & \sim p(w)
\end{align}

whose implied independencies coincide with those implied by the graph shown in Figure \ref{fig:feedforward-scm}.

\end{document}